\documentclass[letterpaper]{siamltex}

\usepackage{amsmath,amssymb,%
graphicx}

\def\ignore#1{}

\newtheorem{example}{Example}

\title{Projective geometry of human motion, with an 
application to injury risk }
\author{H. Laurie\footnotemark[2]
\and R. Penne\footnotemark[3]}

\begin{document}
\maketitle
%\date{}

\renewcommand{\thefootnote}{\fnsymbol{footnote}}

\footnotetext[2]{{\tt henri@pc001a.mth.uct.ac.za} or 
Maths and Applied Maths, UCT, Rondebosch, 
7701, South Africa} 
\footnotetext[3]{{\tt rudi.penne@kdg.be} 
or Industrial Sciences and Technology, Karel de Grote-Hogeschool, Antwerpen}

\renewcommand{\thefootnote}{\arabic{footnote}}

\bibliographystyle{siam}

\pagestyle{myheadings}
\thispagestyle{plain}
\markboth{H. LAURIE AND R. PENNE}{PROJECTIVE GEOMETRY OF HUMAN MOTION}

\begin{abstract}
We give an exposition of 
Pl\"{u}cker vectors for a system of joint axes in projective 3-space. 
We use Pl\"{u}cker vectors to analyse dependencies among joint axes, 
and in particular 
show that two rotational joints rigidly joined by a bar and 
each with 3~degrees of freedom always forms a 5-dimensional system. 
We introduce the concept
of  reduced redundancy in a dependent
set of projective Lines, and argue that reduced redundancy in 
the axes of a body position  
increases injury risk.
We apply this to a simple two-joint model
of bowling in cricket, and show by analysis of some experimental data
that reduced redundancy around ball release is observed in 
some cases.
\end{abstract}


\section{Introduction}

A variety of techniques exist for the mathematical analysis of human 
motion, including techniques that are 
also used in robotics~\cite{potketal01:305,chezetal96:477}. However, to our 
knowledge, nobody has yet employed 
the formalisms and insights of projective geometry, well
known in robotics~\cite{Wh:robot,crai89}.

We are motivated by the analysis of
certain complicated athletic effects achieved by throw-like motions, 
such as a top-spin serve in tennis and 
an away-swinger in cricket. It is clear that the brief interval ending in
the release of the ball is crucial: after release, the ball is in free fall,
except for some aerodynamic and gyroscopic effects. Thus the athlete must
release the ball in a particular state of motion (translational as well 
as rotational).

Several questions arise: by what movements of the joints does a given
athlete achieve a given effect; is there more than one way to achieve a
given effect; do some effects require motions that are inherently more risky
than others, and if so, can this risk be characterized analytically?

Many of these questions can be illuminated by using techniques from the
mathematics of robotics, in which the following are possible: a simple
description which unifies all aspects of the motion of the athlete and the
ball, a representation in which rotation and translation are easily
combined, and a level of generality at which all cases of reduced mobility
can be found (and explicitly calculated).

In this report, we analyse what appears to us as a simple interesting case: 
the motion of a cricketer's arm (much simplified) near the moment of
delivery. We regard the hips as fixed, the torso as rigid, and the 
waist and shoulder as joints,
each of which provides 3~degrees of freedom. Alert readers will notice
that we ignore the elbow, wrist and fingers, as 
well as any contact motion of the ball in the hand prior to release. For our
purposes, it is assumed that the requirements of a given delivery have
prescribed the motion at the centre of the wrist. However, we will see
that in our model in all positions there are only 5~degrees of freedom for
the motion of the ball. Consequently, our system of 6~axes possesses
intrinsic redundancy by design! It can be regarded as the solution of nature
to the human desire to accomplish complicated motions. Indeed, kinematic
redundancy offers an opportunity to distribute stress over many joints
(see also~\cite{potketal01:305}). Our
main result is to prove the existence of special positions where \emph{%
reduced redundancy} occurs: that is, for a given motion the amount of
rotation about one or more joint axes is fixed, while there is some freedom in
distributing motion about the remaining joint axes. These positions should
not be confused with standard kinematic singularities, as no decrease of
mobility is involved; 5 degrees of freedom are always maintained. We
interpret reduced redundancy as a source of injury risk.

The plan of the paper is as follows: first, we describe the use of
Pl\"{u}cker coordinates as a unified framework for computations concerning
rotations and translations in many linked joints. Second, we describe a
simplified model for the motion of an arm in the act of releasing a legal
cricket delivery, a motion associated with risk of overuse
injury~\cite{mcgrfinc96,portetal00:999,lloyetal00:975}. 
Third, we analyze the degrees of freedom for the motion of
the ball in this model, for the general case as well as all special cases;
this includes a full analysis of reduced redundancy. Fourth, we discuss the
analysis from two points of view: the prevention of injury and the forbidden
motions of the wrist. Finally, we show that reduced redundancies indeed 
occur in real bowling actions by analysing data from two bowlers with
an injury history.

\section{Pl\"{u}cker coordinates for human motion}

Human motion is the result of rotations around joint axes, at least, 
infinitesimally in the
first approximation (that is, neglecting the play in the joints
and the deformation of bone, cartilage
and ligament). However, the desired motion of the end effector (in our case,
the cricket ball) will in general have components of both rotation and
translation. In projective geometry, translation can be rendered as rotation
about an axis at infinity. In this view, all motions are rotations, and
Pl\"{u}cker coordinates are merely a convenient way of describing them.

\subsection{Projective Points\label{pro}}

In projective geometry, we identify all points on a line through the origin
in $\mathbb{R}^{4}$ with a projective Point in the corresponding projective
space $\mathbb{P}^{3}$. So the vector 
$\mbox{\bf x}=(\lambda a,\lambda b,\lambda
c,\lambda d)\in\mathbb{R}^{4}$ corresponds to $\mbox{\bf p}\in \mathbb{P}%
^{3}$ for all $\lambda \neq 0$. Such a 4-vector is referred to as a set of
homogeneous coordinates for $\mbox{\bf p}$. By the usual convention, 
the hyperplane $%
H:\quad\! x_{4}=1$ in $\mathbb{R}^{4}$ is considered as (a copy of) affine
3-space. All Points of $\mathbb{P}^{3}$ which correspond to lines
intersecting this hyperplane are called \emph{finite points}, and these are
identified with the affine point of $H$ where they intersect. So for finite
points: 
\begin{equation*}
(a,b,c,d)\sim (a/d,b/d,c/d,1)\sim (a/d,b/d,c/d) 
\end{equation*}
Notice that some lines through the origin in $\mathbb{R}^{4}$ do not
intersect the hyperplane $H$, and therefore some projective Points are not
finite. They are said to lie \emph{at infinity}, and they are represented by
homogeneous coordinates with 0 as fourth coordinate: $(a,b,c,0)$.

Similarly, planes through the origin of $\mathbb{R}^{4}$ correspond to Lines
in $\mathbb{P}^{3}$. If such a plane is parallel to the hyperplane $H$, then
it represents a Line at infinity. Finally, each 3-dimensional subspace of $%
\mathbb{R}^{4}$ is associated with a Plane in $\mathbb{P}^{3}$. The Plane
corresponding to $x_{4}=0$ is the Plane at infinity of $\mathbb{P}^{3}$, and
it contains all Points at infinity.

\subsection{Pl\"{u}cker coordinates\label{plu}}

From an algebraic point of view, a chosen set of homogeneous coordinates for
a Point $\mathbf{p\in }\mathbb{P}^{3}$ represents a vector in the vector
space $V=\mathbb{R}^{4}$. Now we can consider the \emph{exterior algebra }%
built on $V$: 
\begin{equation*}
\wedge V=V^{(0)}\mathbb{\oplus }V^{(1)}\oplus V^{(2)}\oplus V^{(3)}\oplus
V^{(4)} 
\end{equation*}
which enables us to make computations with scalars ($\mathbb{R}=V^{(0)}$),
vectors ($V=V^{(1)}$), but also with more complicated objects called \emph{%
anti-symmetric tensors }, and this in the same framework. The 
\emph{exterior product} $\wedge $ is a bilinear, anti-symmetric operation on 
$\wedge V$, such that for $\mathbf{A}\in V^{(i)}$ and $\mathbf{B}\in V^{(j)}$
we get $\mathbf{A\wedge B}\in V^{(i+j)}$ if $i+j\leq 4$ or\textbf{\ }$%
\mathbf{A}\wedge \mathbf{B=0}$ otherwise (also in the case $i+j\leq 4$ it can
happen that $\mathbf{A}\wedge \mathbf{B=0}$ in $V^{(i+j)}$). 

\begin{example}
The elements in $%
V^{(2)}$ (the so-called 2-tensors) are products $\mathbf{p}\wedge \mathbf{q%
}$ of vectors $\mathbf{p}$ and $\mathbf{q}$ in $V$, or linear combinations
of these. Notice that $\mathbf{p}\wedge \mathbf{q=0}$ in $V^{(2)}$ if $\ 
\mathbf{p}$ and $\mathbf{q}$ represent the same projective Point, due to the
anti-symmetry.
\end{example}

For the reader who is not familiar with the exterior algebra it suffices to
know for our purposes that each tensor can be regarded as just some
vector, and $V^{(i)}$ as a real vector space of dimension $\binom{4}{i}$.
For example, $V^{(2)}\cong \mathbb{R}^{6}$. Furthermore, using the standard
basis of $V$, there is a canonical way to construct a basis for $V^{(i)}$.
The corresponding coordinates arising in this manner for tensors are
called \emph{Pl\"{u}cker coordinates}. 

Let us be more specific in the case of 2-tensors, because they will be needed most
in this article.
If $\mathbf{L}\in
V^{(2)} $ then we have 6 Pl\"{u}cker coordinates for $\mathbf{L}$, by
convention labelled by double-indices: 
\begin{equation*}
\mathbf{L}=(L_{12},L_{13},L_{14},L_{23},L_{24},L_{34}) 
\end{equation*}
In the special case that $\mathbf{L=p\wedge q}$ with $\mathbf{p}%
=(p_{1},p_{2},p_{3},p_{4})$ and $\mathbf{q}=(q_{1},q_{2},q_{3},q_{4})$ there
is an easy rule to obtain the coordinates for $\mathbf{L}$: $%
L_{ij}=p_{i}q_{j}-p_{j}q_{i}$. 
\begin{equation*}
\mbox{\bf p}\wedge \mbox{\bf q}= 
\begin{pmatrix}
p_{1}q_{2}-p_{2}q_{1} \\ 
p_{1}q_{3}-p_{3}q_{1} \\ 
p_{1}q_{4}-p_{4}q_{1} \\ 
p_{2}q_{3}-p_{3}q_{2} \\ 
p_{2}q_{4}-p_{4}q_{2} \\ 
p_{3}q_{4}-p_{4}q_{3}
\end{pmatrix}
\end{equation*}
the elements of which are the $2\times 2$ minors of the matrix $
\begin{pmatrix}
p_{1} & p_{2} & p_{3} & p_{4} \\ 
q_{1} & q_{2} & q_{3} & q_{4}
\end{pmatrix}
$ in lexicographic order. If $\mathbf{p}$ and $\mathbf{q}$ represent
different projective Points then $\mathbf{L=p\wedge q}$ represents the
projective Line through these two Points. Of course, many other 2-tensors
represent the same Line in $\mathbb{P}^{3}$. Indeed, we can use a multiple
of $\mathbf{p}$ or $\mathbf{q}$, without changing the involved projective
Points, or we can even choose another pair of Points on the same Line.
Fortunately, the new 2-tensor $\mathbf{L}^{\prime }\mathbf{=p}^{\prime }%
\mathbf{\wedge q}^{\prime }$ will always be a multiple of $\mathbf{L}$: $%
\mathbf{L}^{\prime }=\alpha \mathbf{L}$. We conclude that the Pl\"{u}cker
coordinates of $\mathbf{L}$ can be considered as a 6-tuple of homogeneous
coordinates for the projective Line represented by $\mathbf{L}$. Notice that
Lines at infinity are characterized by having Pl\"{u}cker coordinates with $%
L_{14}=L_{24}=L_{34}=0$.

Because not every 2-tensor in $V^{(2)}$ can be written as the exterior
product of two vectors in $V$, not every 6-tuple of Pl\"{u}cker coordinates
represents a Line in $\mathbb{P}^{3}$. More precisely, one can prove that $%
(L_{12},L_{13},L_{14},L_{23},L_{24},L_{34})$ corresponds to a projective
Line if and only if it differs from zero and the \emph{Grassmann-Pl\"{u}cker
relation} is satisfied: 
\begin{equation*}
L_{14}L_{23}-L_{24}L_{13}+L_{34}L_{12}=0\qquad \qquad \qquad \text{(GP)} 
\end{equation*}
So, ``most'' 6-tuples in $\mathbb{R}^{6}$ are not the Pl\"{u}cker
coordinates of a projective Line. However, there is an interesting theorem, 
\emph{Poinsot's Central Axis Theorem}, which says that each 2-tensor $%
\mathbf{A}$ not obeying (GP) can be expressed as a sum $\mathbf{A=L+M}$ such
that

\begin{enumerate}
\item  $\mathbf{L}$ corresponds to a finite Line (not at infinity).

\item  $\mathbf{M}$ corresponds to a Line at infinity.

\item  Every affine Plane through $\mathbf{M}$ is perpendicular to $\mathbf{L%
}$.
\end{enumerate}

Let us give one more illustration of Pl\"{u}cker coordinates.
Given is a
2-tensor $\mathbf{A}=(A_{12},A_{13},A_{14},A_{23},A_{24},A_{34})$ and a
vector $\mathbf{p=}(p_{1},p_{2}$\bigskip $,p_{3},p_{4})$. Then $\mathbf{%
P=A\wedge p}$ belongs to the 4-dimensional space $V^{(3)}$: $\mathbf{P}%
=(P_{123},P_{124},P_{134},P_{234})$, with 
\begin{eqnarray*}
P_{123} &=&A_{12}p_{3}-A_{13}p_{2}+A_{23}p_{1} \\
P_{124} &=&A_{12}p_{4}-A_{14}p_{2}+A_{24}p_{1} \\
P_{134} &=&A_{13}p_{4}-A_{14}p_{3}+A_{34}p_{1} \\
P_{234} &=&A_{23}p_{4}-A_{24}p_{3}+A_{34}p_{2}
\end{eqnarray*}
In particular, if $\mathbf{A}$ represents a projective Line, which moreover
does not contain the projective Point represented by $\mathbf{p}$, then $%
\mathbf{P}$ represents the projective Plane determined by this Line and this
Point. If the Point lies on the Line, then $\mathbf{P=0}$. In any case, if $%
\ $a 3-tensor differs from zero, it will represent a Plane in $\mathbb{P}%
^{3}$. Furthermore, it is the Plane at infinity iff $%
P_{124}=P_{134}=P_{234}=0$. On the other hand, if $\mathbf{P}\in V^{(3)}$
represents a finite Plane, the vector $(P_{234},-P_{134},P_{124})\in 
\mathbb{R}^{3}$ is perpendicular to the associated affine plane.

For a good introduction to Pl\"{u}cker coordinates and anti-symmetric tensors, 
including the formal definitions,
we refer to \cite{Wh:robot}.

\subsection{Dependencies among lines\label{depend}}

A set of Lines in $\mathbb{P}^{3}$, finite or at infinity, is called
independent (resp., dependent), if the corresponding 2-tensors are
linearly independent (resp., dependent) in $V^{(2)}$, or equivalently, if the
corresponding Pl\"{u}cker coordinates are 
linearly independent (resp., dependent) in $%
\mathbb{R}^{6}$. These concepts are defined in algebraic terms, nevertheless
the possible dependencies among projective Lines have a transparent
geometric characterization. We refer to \cite{Da} for a complete description of
this. We only quote those situations that will be relevant for our analysis.

\begin{itemize}
\item  Two Lines can only be dependent when they coincide.

\item  Three Lines are dependent if and only if they lie in the same Plane 
\textbf{and} go through the same Point.

\item  Four Lines are dependent if and only if at least one of the following
cases occur:
\begin{enumerate}
\item Three of the four Lines are dependent.
\item The four Lines lie in the same Plane.
\item The four Lines go through the same Point.
\item Two of the Lines lie in a Plane $\alpha$, intersecting in Point $p$, and
the remaining two Lines lie in a Plane $\beta$, intersecting in Point $q$, such
that the Planes $\alpha$ and $\beta$ meet in the Line $pq$.
\item The four Lines belong to the same system of rulers on a quadratic surface.
\end{enumerate}
\end{itemize}

\bigskip

In particular, if we are given two parallel Lines (intersecting at
infinity), then a linear combination of their Pl\"{u}cker coordinates will
always represent a Line in the unique plane through the given Lines, and
which either lies at infinity, or which is parallel to the given Lines. Four
parallel Lines in 3-space are always dependent.

\begin{example}
As an illustration, let us consider a situation of 4 Lines, with $\mathbf{L}%
_{1}\mathbf{,L}_{2}\mathbf{,L}_{3}$ concurrent (but not coplanar) through
Point $\mathbf{p}$, and $\mathbf{L}_{4}$ not containing $\mathbf{p}$.
Clearly, these 4 Lines are independent. By taking linear combinations of the
3 concurrent Lines we can generate any Line $\mathbf{L}$ through $\mathbf{p}$%
. Furthermore, if $\mathbf{L}$ happens to intersect $\mathbf{L}_{4}$ (in $%
\mathbf{q}$ say), then linear combinations of $\mathbf{L}$ and $\mathbf{L}%
_{4}$ generate Lines through $\mathbf{q}$, lying in the Plane determined by $%
\mathbf{L}$ and $\mathbf{L}_{4}$. If $\mathbf{L}$ and $\mathbf{L}_{4}$ do
not intersect each other, then we cannot obtain new Lines by combining them
(violation of (GP)!). We conclude that the Lines which depend on $\mathbf{L}%
_{1}\mathbf{,L}_{2}\mathbf{,L}_{3}\mathbf{,L}_{4}$, are exactly those that
contain $\mathbf{p}$ or that lie in the Plane through $\mathbf{L}_{4}$ and $%
\mathbf{p}$.
\end{example}

Next, we will elaborate a special case which will be important for the 
applications in this paper.

\begin{theorem}
\label{rank5}
{\Large \ }Let $\mathbf{W}_{1}\mathbf{,W}_{2}\mathbf{,W}_{3}$ and $\mathbf{S}%
_{1}\mathbf{,S}_{2}\mathbf{,S}_{3}$ be two triples of concurrent Lines in $%
\mathbb{P}^{3}$, through different Points $\mathbf{w}$ and $\mathbf{s}$. 
Assume moreover that $\mathbf{W}_{1}\mathbf{,W}_{2}%
\mathbf{,W}_{3}$ are not coplanar, neither are $\mathbf{S}_{1}\mathbf{,S}_{2}%
\mathbf{,S}_{3}$. Then the set $\{\mathbf{W}_{1}\mathbf{,W}_{2}\mathbf{,W}%
_{3},\mathbf{S}_{1}\mathbf{,S}_{2}\mathbf{,S}_{3}\}$ always has rank 5. Or
more explicitly, these 6 Lines are always dependent, but always contain a
subset of 5 Lines which is independent.
\end{theorem}

\begin{proof}
Choose a Pl\"{u}cker vector $\mathbf{P}$ to represent the Line $\mathbf{sw}$%
. By abuse of notation, we let $\mathbf{W}_{1},\ldots ,\mathbf{S}_{3}$ stand
for the Pl\"{u}cker vectors of the corresponding lines as well. Then there 
exist linear combinations 
\begin{eqnarray*}
\mathbf{P} &=&\mathbf{\alpha }_{1}\mathbf{W}_{1}\mathbf{+\alpha }_{2}\mathbf{%
W}_{2}\mathbf{+\alpha }_{3}\mathbf{W}_{3} \\
\mathbf{P} &=&\mathbf{\beta }_{1}\mathbf{S}_{1}\mathbf{+\beta }_{2}\mathbf{S}%
_{2}\mathbf{+\beta }_{3}\mathbf{S}_{3}
\end{eqnarray*}
which gives rise to the claimed dependency: 
\begin{equation*}
\mathbf{\alpha }_{1}\mathbf{W}_{1}\mathbf{+\alpha }_{2}\mathbf{W}_{2}\mathbf{%
+\alpha }_{3}\mathbf{W}_{3}-\mathbf{\beta }_{1}\mathbf{S}_{1}-\mathbf{\beta }%
_{2}\mathbf{S}_{2}-\mathbf{\beta }_{3}\mathbf{S}_{3}=\mathbf{0}
\end{equation*}
Next, we observe that at least one of $\{\mathbf{S}_{1}\mathbf{,S}_{2}%
\mathbf{,S}_{3}\}$ does not pass through $\mathbf{w}$, say $\mathbf{S}_{1}$.
From the example above we learn that $\{\mathbf{W}_{1}\mathbf{,W}_{2}\mathbf{%
,W}_{3}\mathbf{,S}_{1}\}$ is a set of independent Lines, and moreover, the
only Lines which are dependent on these 4 Lines, are Lines through $\mathbf{w%
}$, or Lines in the Plane determined by $\mathbf{S}_{1}$ and $\mathbf{w}$.
Because the triple $\{\mathbf{S}_{1}\mathbf{,S}_{2}\mathbf{,S}_{3}\}$ is
assumed to be non-planar, it is impossible that both $\mathbf{S}_{2}$ and $%
\mathbf{S}_{3}$ depend on $\{\mathbf{W}_{1}\mathbf{,W}_{2}\mathbf{,W}_{3}%
\mathbf{,S}_{1}\}$, which completes the proof.
\end{proof}

\bigskip

\subsection{Describing kinematics by Pl\"{u}cker coordinates}

For a more extended exposition of the material presented in this 
paragraph, we refer to \cite{Wh:robot} and \cite{CrWh}.
Consider a motion of a rigid body $B$ in 3-space. Then, every point $p$ of $%
B $ traces a path, $p=p(t)$. If the motion is sufficiently \emph{smooth}
from a mathematical point of view, we can compute the derivative at a
certain time $t_{0}$, giving us the infinitesimal motion of $B$ at $t=t_{0}$%
. This results in a velocity vector $v_{p}=%
%\overset{\cdot }{p} 
\dot{p}(t_{0})$
for
every point $p$ of $B$. The rigidity of $B$ can be translated into the
statement that for every pair of its points $\{p,q\}$ the distance between these
points must remain constant during the motion: 
\begin{equation*}
||p(t)-q(t)||^{2}=\text{constant} 
\end{equation*}
or, infinitesimally (\emph{preserved distance property}): 
\begin{equation*}
(v_{p}-v_{q})\cdot (p-q)=0\qquad \qquad \qquad \text{(PDP)} 
\end{equation*}
From now on, when we use the term ``motion'', we always mean an
infinitesimal rigid motion: the assignment of a velocity vector to every
point of $B$, such that (PDP) is satisfied. 
So, we
associate a vector $v_{p}$ to every point $p$ of $B$, taking (PDP) into
account.

One important example of such a motion is a spatial rotation about the
origin. Here, there is always a line $A$ involved, the so-called axis of
rotation, containing the origin. Points on $A$ remain fixed (zero velocity
vector), but for other points $p$ the velocity $v_{p}$ is perpendicular to
the plane determined by $A$ and $p$. As a matter of fact, the rotation is
specified by a vector $\omega $ along $A$, such that $v_{p}=\omega \times p$
(vector cross product). The length of $\omega $ is called the angular velocity,
and, together with the distance of $p$ from the axis $A$, it determines the
length of $v_{p}$.

Another fundamental motion is a translation along a given vector $v$. Here,
we have a constant velocity: for every point $p$ we put $v_{p}=v$. 

A crucial theorem says that every rigid motion is the composition of
rotations and translations, or infinitesimally, the velocity vectors can be
written as the sum of rotation velocities and/or translation velocities.

Consider a rotation about some axis $A$, not necessarily containing the
origin. If we embed affine 3-space into $\mathbb{P}^{3}$, as described in
Section \ref{pro}, then we can associate with $A$ a projective Line $\mathbf{%
A}$, and hence a Pl\"{u}cker vector $\mathbf{P}_{A}$. For each point $p$ in $%
\mathbb{R}^{3}$ we choose the standard homogeneous coordinates for the
associated projective Point $\mathbf{p}$ (having $p_{4}=1$). Now we can
define the ``motion of $\mathbf{p}$'' as the following 3-tensor: 
\begin{equation*}
M(\mathbf{p)=M=P}_{A}\wedge \mathbf{p}\in V^{(3)} 
\end{equation*}
To see that this makes sense, consider a vector 
$\mathbf{M%
}=(M_{123},M_{124},M_{134},M_{234})$ of Pl\"{u}cker coordinates. 
This determines the vector $%
v_{p}=(M_{234},-M_{134},M_{124})\in \mathbb{R}^{3}$, which is zero if $p\in A
$, or else, it is perpendicular to the plane determined by $p$ and $A$. And
indeed, as one can prove that (PDP) holds for these vectors, they represent
a rotation about axis $A$. The unused coordinate $M_{123}$ in $M(\mathbf{p})$
is determined by the fact that this 3-tensor corresponds to a plane
through $\mathbf{p}$. Of course, the magnitude of the vectors $v_{p}$
depends on the chosen Pl\"{u}cker coordinates $\mathbf{P}_{A}$ for $\mathbf{A%
}$, but then again there are an infinite number
of possible rotations about axis $%
A$ in $\mathbb{R}^{3}$. One can say that the magnitude of the chosen
Pl\"{u}cker vector accounts for the involved angular velocity. We conclude
that the 2-tensor $\mathbf{P}_{A}$ encodes both the rotation axis $A$ and
the angular velocity. Therefore, it is called the \emph{center} of the
motion. Taking a multiple of $\mathbf{P}_{A}$ does not change the axis, but
only the angular velocity. If you are interested in the velocity of a
specific point $p$ under this motion, just perform the exterior product $%
\mathbf{P}_{A}\mathbf{\wedge p}$, using standard homogeneous coordinates 
for $\mathbf{p}$.

Now we have put spatial rotations in the setting of projective geometry, we
can extend the notion of rotation axis. Indeed, we can take $\mathbf{A}$ to
be a Line at infinity, so, if $\mathbf{P=P}_{A}$ then $\mathbf{P}_{14}%
\mathbf{=P}_{24}\mathbf{=P}_{34}\mathbf{=0}$. If we copy the previous
computations for some point $p$, we observe, surprisingly, that the last
three Pl\"{u}cker coordinates of $M(\mathbf{p)}$ do not depend on $p$. So,
we see that $v_{p}$ is a constant vector if we perform a rotation about an
axis at infinity, which must be a translation! More precisely, $v_{p}\mathbf{%
=(P}_{23}\mathbf{,-P}_{13}\mathbf{,P}_{12}\mathbf{)}$, a vector which is
perpendicular to any plane in $\mathbb{R}^{3}$ whose projective extension
contains the given axis at infinity $\mathbf{A}$. For the sake of
uniformity, we will call the 2-tensor $\mathbf{P}_{A}$ again the center of
the motion, and the 3-tensor $M(\mathbf{p)}$ the motion itself of the
point $p$.

Our arguments will directly take place in $\mathbb{P}^{3}$ or $\wedge 
\mathbb{R}^{4}$, but readers who like to switch to affine space now and then
should remember: 
\begin{eqnarray*}
p=(p_{1},p_{2},p_{3})\longrightarrow &&\mathbf{p=(}p_{1},p_{2},p_{3},1) \\
v_{p}=(M_{234},-M_{134},M_{124})\longleftarrow &&M(\mathbf{p)=}%
(M_{123},M_{124},M_{134},M_{234})
\end{eqnarray*}
In this setting, the zero tensor in $V^{(3)}$ corresponds to the zero velocity.

As mentioned before, composing two motions comes down to adding the velocity
vectors in each point $p$. Let the corresponding centers of motion be
denoted by $\mathbf{C}_{1}$ and $\mathbf{C}_{2}$, Pl\"{u}cker vectors in $%
\mathbb{R}^{6}$. Then the resulting motion of $p$ equals 
$$\mathbf{C}_{1}\wedge \mathbf{p+C}_{2}\wedge%
\mathbf{p=(C}_{1}+\mathbf{C}_{2})\wedge \mathbf{p},$$%
due to a basic property of the exterior product. Now we can
consider $\mathbf{C=C}_{1}\mathbf{+C}_{2}$ to be the center of the composite
motion. This means that every Pl\"{u}cker vector $\mathbf{P}$ in $\mathbb{R}%
^{6}$ can play the part of a center of some motion. More precisely, if $%
\mathbf{P}$ represents a projective Line (satisfying (GP)) then it gives
rise to a rotation (finite line) or a translation (line at infinity),
otherwise it is the center of a composition of rotations and translations.
As a consequence of Poinsot's Central Axis Theorem (Section \ref{plu}) we
can be even more specific in the latter case. To this end, we define a \emph{%
screw motion} as the composition of a rotation (infinitesimal, of course) 
about some axis, and a translation (ditto) along the
same axis.

\begin{center}
\textbf{If a motion is not a pure translation or rotation then it is a
screw motion.}
\end{center}

From now on, Pl\"{u}cker coordinates of 2-tensors (the space $\mathbb{R}%
^{6}$) are interpreted as centers of infinitesimal rigid motions.

\section{A simple model for bowling a cricket delivery}

\bigskip 
Biomechanical models for cricket motions are not that rare, but few exist for 
bowling~\cite{burnetal98:574,lloyetal00:975}.
For our purposes, a model simpler than either of these will suffice.
We make the following assumptions about motion just
prior to delivery:

1. There is no rotation in the elbow (as is required in a legal 
delivery).

2. There is no rotation in the wrist, and the state of motion of the ball
upon release prescribes the motion of the so-called tool centre (a term from
robotics), which we take to be the wrist.

3. The spine is taken as rigid, but free to rotate as if its base attached
to the pelvis in a ball joint (i.e.\ we ignore deformation of the torso), 
and the shoulder is rigidly joined to the spine.

4. The joint axes of both  joints pass through the centre of the joint.

For greater realism, one might add more joints, for example it is
known that the shoulder does rotate relative to the
torso~\cite{ellietal02:507}, 
and the ball might leave the hand in a contact motion. This is not
conceptually  
difficult, but is computationally and experimentally  
challenging. The same applies to relaxing assumption~4, to allow
non-coincident joint axes. Still more challenging would be the direct 
modelling of
muscle groups (as in~\cite{eberetal99:1,vdboscha93:95}), 
as this would increase the number of axes of rotation
substantially, and one might be hard put to identify an axis of rotation for
every muscle group, particularly those with attachments over more than one
joint.

\subsection{Introducing the joint axes of our model}

With assumptions~1 to~4, the system reduces to two joints, which we call the
waist ($w$) and the shoulder ($s$). Although in general $w$ may be in
motion, there is no loss of generality if we place $w$ at the fixed origin,
and identify its joint axes with axes of a reference coordinate system $XYZ$.
They are interpreted as follows: for a person standing,
$X$ points horizontally forward, $Y$ horizontally points to the
left, and $Z$ points vertically along the spine, in our case upwards. 

\begin{figure}
\includegraphics[height=60mm]{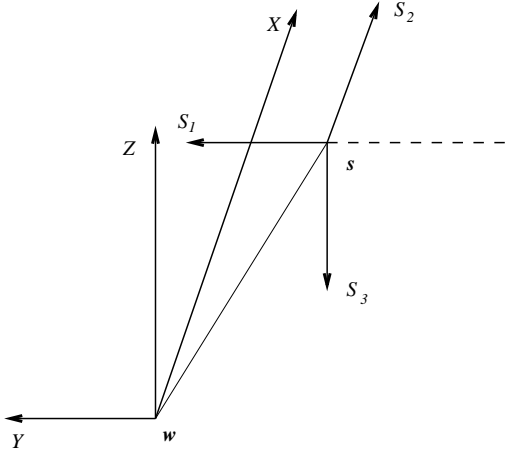}
\caption{Configuration of joint axes in our simplified model of a right-handed
cricket bowler, facing away from the reader.
{\bf w} is the waist joint and {\bf s} is the shoulder joint. Distances are
normalised so that the shoulder is at $(0, -1, 1)$ in the waist joint
axes. The dashed line corresponds to the arm in standard position:
horizontal, palm down. Note that the shoulder axes move with the arm,
so that $S_3$ is always pointing in the same direction as the palm.}
\label{genconfig}
\end{figure}

We choose units of length so that the right shoulder joint $s$ is at
$(0,-1,1)$ in the system; since the torso is rigid it stays there. The
three joint axes through $s$ follow the usual convention: 
we choose $S_{1}$ as the axis that passes through shoulder and elbow,
in the direction of the shoulder, and $S_{2},S_{3}$
perpendicular to each other and to $S_{1}$, so that when the arm is 
extended sideways horizontally wrist down, $S_2$ points forwards and 
$S_3$ points downwards.  This means that the $S_1S_2S_3$ system moves
with the arm, and in particular that $S_3$ is always perpendicular to
the palm. The general configuration is illustrated in
Figure~\ref{genconfig}.

We note that some of our results below depend on the choice of
shoulder axes. In particular, we find a case where rotation about the
$s_3$ axis plays a significant role in predicting injury risk. This
would be indefensible if all we knew of the shoulder joint was that it
had 3~degrees of rotational freedom, because our result would
disappear under many other apparently equivalent choice of
axes.

However, we do know more about how the shoulder moves and about the
motion of the arm of a fast bowler near the point of release. 
In that context, the $s_1s_2s_3$ system as described above is
preferred and has intrinsic intepretation  for two reasons. Firstly,
$s_1$ is an anatomically intrinsic axis in all rotations 
of the shoulder, because of the role of the rotator cuff, which are
the only muscles that cause rotation around $s_1$.  Secondly, during
the final phase of the delivery of a fast ball in cricket, the
bowler's arm moves in a plane. Near the moment of release, the direction of
$s_3$ is tangential to this motion (since these bowlers aim for high
speed deliveries), so the plane of motion is the $s_1s_3$ plane, and
in that plane the motion is a pure rotation around the $s_2$ axis. By
orthogonality to both the $s_1$ and $s_2$ axes, the $s_3$ axis is also
intrinsic.  

\subsection{Pl\"{u}cker coordinates of the 6 joint axes\label{choices}}

The positions of the waist and the shoulder are given by 
\begin{equation*}
w = \begin{pmatrix}
0& 0 & 0 & 1 \end{pmatrix} \qquad \text{and}\qquad 
s = \begin{pmatrix}0 & -1 & 1 & 1 \end{pmatrix}. 
\end{equation*}

The directions of the joint axes are 
\begin{eqnarray*}
W_{1} &=& 
\begin{pmatrix}
1 & 0 & 0 & 0
\end{pmatrix}
\\
W_{2} &=& 
\begin{pmatrix}
0 & 1 & 0 & 0
\end{pmatrix}
\\
W_{3} &=& 
\begin{pmatrix}
0 & 0 & 1 & 0
\end{pmatrix}
\\
S_{1} &=& 
\begin{pmatrix}
a_{1} & b_{1} & c_{1} & 0
\end{pmatrix}
\\
S_{2} &=& 
\begin{pmatrix}
a_{2} & b_{2} & c_{2} & 0
\end{pmatrix}
\\
S_{3} &=& 
\begin{pmatrix}
a_{3} & b_{3} & c_{3} & 0
\end{pmatrix}
\end{eqnarray*}
The six centers
 of rotation are then 
\begin{eqnarray*}
%\mbox{\bf p}
P_{1}=W_{1}\wedge w= 
\begin{pmatrix}
1 & 0 & 0 & 0 \\ 
0 & 0 & 0 & 1
\end{pmatrix}
&=& 
\begin{pmatrix}
0 & 0 & 1 & 0 & 0 & 0
\end{pmatrix}
\\
%\mbox{\bf p}
P_{2}=W_{2}\wedge w= 
\begin{pmatrix}
0 & 1 & 0 & 0 \\ 
0 & 0 & 0 & 1
\end{pmatrix}
&=& 
\begin{pmatrix}
0 & 0 & 0 & 0 & 1 & 0
\end{pmatrix}
\\
%\mbox{\bf p}
P_{3}=W_{3}\wedge w= 
\begin{pmatrix}
0 & 0 & 1 & 0 \\ 
0 & 0 & 0 & 1
\end{pmatrix}
&=& 
\begin{pmatrix}
0 & 0 & 0 & 0 & 0 & 1
\end{pmatrix}
\\
%\mbox{\bf p}
P_{4}=S_{1}\wedge s= 
\begin{pmatrix}
a_{1} & b_{1} & c_{1} & 0 \\ 
0 & -1 & 1 & 1
\end{pmatrix}
&=& 
\begin{pmatrix}
-a_{1} & a_{1} & a_{1} & b_{1}+c_{1} & b_{1} & c_{1}
\end{pmatrix}
\\
%\mbox{\bf p}
P_{5}=S_{2}\wedge s= 
\begin{pmatrix}
a_{2} & b_{2} & c_{2} & 0 \\ 
0 & -1 & 1 & 1
\end{pmatrix}
&=& 
\begin{pmatrix}
-a_{2} & a_{2} & a_{2} & b_{2}+c_{2} & b_{2} & c_{2}
\end{pmatrix}
\\
%\mbox{\bf p}
P_{6}=S_{3}\wedge s= 
\begin{pmatrix}
a_{3} & b_{3} & c_{3} & 0 \\ 
0 & -1 & 1 & 1
\end{pmatrix}
&=& 
\begin{pmatrix}
-a_{3} & a_{3} & a_{3} & b_{3}+c_{3} & b_{3} & c_{3}
\end{pmatrix}
\end{eqnarray*}

We collect these in the columns of the motion matrix $M$: 
\begin{equation*}
M=
\begin{pmatrix}
0 & 0 & 0 & -a_{1} & -a_{2} & -a_{3} \\ 
0 & 0 & 0 & a_{1} & a_{2} & a_{3} \\ 
1 & 0 & 0 & a_{1} & a_{2} & a_{3} \\ 
0 & 0 & 0 & b_{1}+c_{1} & b_{2}+c_{2} & b_{3}+c_{3} \\ 
0 & 1 & 0 & b_{1} & b_{2} & b_{3} \\ 
0 & 0 & 1 & c_{1} & c_{2} & c_{3}
\end{pmatrix}
\end{equation*}
All the information regarding configurations of the joints and possible
motions can be found by analysing $M$. More precisely, infinitesimally, the
motion of the wrist is a composition of a rotation about $w$
and a rotation about $s$ (in our model). So, the center of this motion is a
linear combination of the six Pl\"{u}cker coordinates which we assigned to
the six given axes. This motivates us to define the column space of the
matrix $M$ to be the\emph{\ motion space }(of the wrist in the given position
of the human body), $\mathbb{MS}$. Recall from Theorem \ref{rank5} that the
matrix $M$ always has rank equal to 5, implying a constant dimension of 5
for the motion space. Notice that we never obtain the full $\mathbb{R}^{6}$
as motion space in our model; this would require including futher 
rotations in our model, such as 
about the elbow or the wrist.

\subsection{Possible motions under the model}

Suppose the human body (in particular, the torso and the bowling arm) is in
a certain position. If one intends to propel the ball in some specific way,
then this is accomplished by performing an infinitesimal motion with the
hand. In our model, the only way to realize a hand motion is by means of
rotations about the waist (3 joint axes) and/or about the shoulder (3 joint
axes). Every (infinitesimal) rotation about one of these 6 axes is given by
an appropriate multiple of the corresponding Pl\"{u}cker vector. We conclude
that the motion of the ball is controlled by a 2-tensor which is a linear
combination of the 6 Pl\"{u}cker vectors of our model, that is, it belongs
to the column space of the matrix $M$ ($\mathbb{MS}$). 
In particular, a linear combination which gives the zero 2-tensor corresponds
to not moving at all (the zero center of motion).

Clearly, the first two rows of $M$ are equal in magnitude but opposite
in sign. This implies that every
possible motion is represented by a 2-tensor with opposite Pl\"{u}cker
coordinates in the first two places:
\begin{equation*}
\mathbf{B}=(-a,a,b,c,d,e)
\end{equation*}
or equivalently, a possible motion is a point of $\mathbb{R}^{6}$ in the
hyperplane $\mathbb{H}:p_{12}=-p_{13}$, so $\mathbb{MS\subset H}$.
Furthermore, since both spaces have dimension 5, we can state that $\mathbb{%
MS=H}$.

\begin{example}
Try to perform a \textbf{pure} translation with your hand along the $Z$-axis
(the direction of the spine) by only using the waist joint and the shoulder
joint. You will not succeed! The algebraic proof for this goes as follows.
Each translation along $Z$ is represented by a set of Pl\"{u}cker
coordinates of the line at infinity of the $XY$-plane. This means that it is
a multiple of
\begin{equation*}
(1,0,0,0)\wedge (0,1,0,0)=(1,0,0,0,0,0)
\end{equation*}
which is not a possible motion, because it does not belong to $\mathbb{H}$.
\end{example}

\begin{example}
In an analogous fashion we see that a \textbf{pure} translation along the $Y$%
-axis is not possible. Indeed, such a translation is always represented by a
multiple of $(0,1,0,0,0,0)$, the Pl\"{u}cker vector for the line at infinity
of the $XZ$-plane.
\end{example}

\begin{example}
However, a translation along the $X$-axis appears to be possible (this is
the direction perpendicular to the plane of the torso; fortunately for 
cricketers, this direction is the one they want the ball to go). Indeed, the
corresponding 2-tensor is a multiple of $(0,0,1,0,0,0)$, the line at
infinity of $YZ$, hence it belongs to $\mathbb{MS}$. But how can this be
accomplished in practice? Let $L$ be the line through $s$ and parallel to $Y$%
. Because the Pl\"{u}cker vector of $L$ is a linear combination of the Pl%
\"{u}cker vectors of $S_{1}$, $S_{2}$ and $S_{3}$, any rotation about $L$
can be realized. Notice that $L$ lies in the $YZ$-plane, as so does the
shoulder joint $s$ in our model. Because $Y$ and $L$ intersect at infinity,
an appropriate linear combination of their Pl\"{u}cker vectors yields the
line at infinity of $YZ$. We conclude that a pure translation along $X$ can
be realized by composing a rotation about $Y$ and a rotation about $L$.
\end{example}

Forbidden motions are interesting for two reasons. Firstly, they are a
simple way of describing what is possible. Second, they have an associated
injury risk: attempting forbidden motion will introduce extremely large
stresses, and coming close to forbidden motion (in the sense of a path
through the motion space) may also require large stresses, a well-known
phenomenon in robotics~\cite{crai89}.

\subsection{Critical positions of the human body}

In our model, the possible motions are supported by six joint axes, each
with a natural physical interpretation. Giving the spatial
positions of these six axes determines what we will call the ``position of
the human body''. In the previous sections we explained that our analysis of
cricket bowling comes down to exploring the linear relations between the 
Pl\"{u}cker coordinates of these six axes. To simplify computations we are
entitled to make the fixed coordinate frame coincide with the three waist
axes, and that is what we did in Section~\ref{choices}, yielding the simple
structure of matrix $M$. So, in fact, by describing a body position we will
mean the specification of the relative position of the shoulder axes with
respect to the waist axes.
It needs 3 parameters to be specified in order to fix the orientation of 
$S_1S_2S_3$
relative to $W_1W_2W_3$ ($=XYZ$), for example the 3 Euler angles.
Thus the very limited positions of the
human body relevant to this paper can be regarded as points in a 
3-dimensional space {\bf Pos}.

\subsection{Redundancy and supports}

As a consequence of Theorem~\ref{rank5} we know that, in each position of the
body, our six joint axes span a 5-dimensional space ($\mathbb{MS}$). We say
that our kinematic system has a \emph{generic redundancy}. Further, still in
each position, basic linear algebra teaches us that we have a 1-dimensional
 space of linear dependencies between our six 2-tensors ($6-5=1$).  
 Redundancy in a model for human motion is also treated
 in~\cite{potketal01:305},  
 where the emphasis is also on the potential for fatigue management
but the operational definition and mathematical treatment  are different.

\begin{definition}
The \emph{support }of a linear dependency amongst a set $A$ of vectors is
the subset of $A$ consisting of exactly those vectors with non-zero
coefficient in this dependency.
\end{definition}

In a given position of the body, each (non-trivial) linear dependency of
the six joint axes is a multiple of every other one. So, we can define merely
the ``support of a body position'', without specifying the linear
dependency. Notice that, whatever position we are in, we always use the
same notation for our six joints axes, hence the support can always be
considered as a subset of $J=\{X,Y,Z,S_{1},S_{2},S_{3}\}$. This can be
mathematically encoded in a map:
\begin{equation*}
\mathbf{supp: }\mathbf{Pos}\rightarrow 2^{J}:p\mapsto \mathbf{supp}(p)
\end{equation*}
Before proceeding, let us explain the relevance of the this concept. Suppose
the body is in some position $p$. Let $\mathbf{M}%
=(M_{12},M_{13},M_{14},M_{23},M_{24},M_{34})$ be the motion in $\mathbb{MS}$
that we want to perform. This is achieved by finding appropriate
coefficients (angular velocities):
\begin{equation*}
\mathbf{M}=\alpha \mathbf{X}+\beta \mathbf{Y}+\gamma \mathbf{Z}+\sigma _{1}%
\mathbf{S}_{1}+\sigma _{2}\mathbf{S}_{2}+\sigma _{3}\mathbf{S}_{3}
\end{equation*}
where the bold font reminds us of the fact that we switched to Pl\"{u}cker
vectors (or 2-tensors). Now suppose that $\mathbf{supp}(p)=\{Y,Z,S_{3}\}$,
corresponding to the following relation:
\begin{equation*}
\lambda \mathbf{Y+}\mu \mathbf{Z}+\nu \mathbf{S}_{3}=\mathbf{0}
\end{equation*}
with nonzero coefficients $\lambda ,\mu ,\nu $. Then we can realize the same
motion $\mathbf{M}$ as 
\begin{equation*}
\mathbf{M}=\alpha \mathbf{X}+(\beta +k\lambda )\mathbf{Y}+(\gamma +k\mu )%
\mathbf{Z}+\sigma _{1}\mathbf{S}_{1}+\sigma _{2}\mathbf{S}_{2}+(\sigma
_{3}+k\nu )\mathbf{S}_{3}
\end{equation*}
with $k$ an arbitrary constant. This means that the efforts done by $Y$, $Z$
and $S_{3}$ can be traded among each other, while the contributions by $X$, $%
S_{1}$ and $S_{2}$ are given by fixed coefficients with no chance for
compensation. From this we learn two important things:

%\begin{description}
%\item 
\begin{enumerate}
\item  The concept of redundancy of joint axes is inherent to the human
body. It is the solution supplied by nature to distribute the necessary
efforts among the several joints for achieving a certain motion.
\item  Positions in which the human body has abundant support are less
strenuous than positions with limited support.
\end{enumerate}
%\end{description}

\subsection{Critical positions}

Now we arrive at the core of this paper. We will classify the possible
supports in our model. A position of the human body is called \emph{critical%
} if the support is smaller than expected, that is, smaller than in generic
positions. We say that a critical position suffers from \emph{
redundancy with reduced support} or shortly, \emph{reduced redundancy}. 
Our first observation says that the required work for joint
axis $X$ can never be compensated by one of the other five axes.

\begin{theorem}
For each position $p\in \mathbf{Pos}$ we have that $X\notin \mathbf{supp}%
(p)$.
\end{theorem}

\begin{proof}
Since the shoulder joint $s$ is assumed to lie in the $YZ$-plane, the Line $%
\mathbf{L}=sw$ is a linear combination of $\mathbf{Y}$ and $\mathbf{Z}$. And
of course, $\mathbf{L}$ is a linear combination of $\mathbf{S}_{1}\mathbf{,S}%
_{2}\mathbf{,S}_{3}$, hence the set $\{\mathbf{S_{1},S_{2},S_{3},Y,Z}\}$ is
dependent. Because the motion space $\mathbb{MS}$ has dimension 5 in every
position, $\mathbf{X}$ cannot be a linear combination of $\mathbf{S}_{1}%
\mathbf{,S}_{2}\mathbf{,S}_{3},\mathbf{Y,Z}$, and hence, it does not belong
to the support.
\end{proof}

\begin{theorem}
Let $p$ be a position of the human body. We distinguish three cases for the
Line $\mathbf{L}=sw$.

\begin{enumerate}
\item  The Line $\mathbf{L}$ is not contained in a plane determined by
  any two Lines of $%
\{\mathbf{S_{1},S_{2},S_{3}}\}$. In this case
\begin{equation*}
\mathbf{supp}(p)=\{\mathbf{S_{1},S_{2},S_{3},Y,Z}\}
\end{equation*}

\item  The Line $\mathbf{L}$ does not coincide with a line of 
$\{\mathbf{S_{1},S_{2},S_{3}}\}$, 
but it lies in the plane generated by two of them ($\mathbf{L}\in 
\mathbf{S_{i}S_{j}}$). Then
\begin{equation*}
\mathbf{supp}(p)=\{\mathbf{S_{i},S_{j},Y,Z}\}
\end{equation*}

\item  The Line $\mathbf{L}$ coincides with one of 
$\{\mathbf{S_{1},S_{2},S_{3}}\}$ (i.e.\  $\mathbf{L}=\mathbf{S_{i}}$). Then
\begin{equation*}
\mathbf{supp}(p)=\{\mathbf{S_{i},Y,Z}\}
\end{equation*}

\end{enumerate}
\end{theorem}

\begin{proof}
The claims are an immediate consequence of what is said in Section~\ref
{depend}.

In case 3, if $\mathbf{L}=\mathbf{S_{i}}$ then the Lines
$\mathbf{Y,Z,S_{i}}$ are concurrent and coPlanar, and so they are
dependent. The support cannot be smaller, because this would mean that
at least two of these lines coincide. 

In case 2, either Lines $\mathbf{S_{i},S_{j},Y,Z}$ are coPlanar, or the pairs
$\mathbf{\{Y,Z\}}$ and $\mathbf{\{S_i,S_j\}}$ 
determine two Planes that meet in the line
$\mathbf{sw}$ through their intersections. In both cases, the four Lines are
dependent.
Furthermore, no three of them are concurrent, implying that the support is
not smaller.

In case 1, we can rule out the five possibilities for the dependency of
four lines (Section 2.3). We refer to Theorem 3.2 for the claim that $%
\{\mathbf{S_{1},S_{2},S_{3},Y,Z}\}$ 
is a dependent set. 
\end{proof}

\noindent{\bf Remark.}
The cases 2 and 3 of the previous theorem correspond to the critical
positions of our model.

\section{Reduced redundancy as injury risk}

It is known that high levels of fitness are attained in many
cricketers~\cite{noakdura00:919}; nevertheless, injuries are fairly
common~\cite{learwhit00:145} and fatigue may play a signficant
role~\cite{elli00:983}. This is not the place to review the
mechanisms of overuse injury (the interested reader is referred to
\cite{vmec92:320} as a starting point). We adopt the common
perspective that overuse injuries start as micro-injuries such as bruised
bone and micro-torn ligament. We suggest that overuse
is more likely in situations of reduced redundancy. In such cases, no
compensation that reduces the strain on a micro-injured site is
possible. The subject, in repeating the action, is condemned to
repeating, at the same intensity, a motion that already caused a
micro-injury. By contrast, the ability to achieve a  
desired motion with a range of different joint rotations amounts to
having the option of avoiding a motion that has caused a micro-injury.
The probability of overuse injury should decrease, hence redundancy
should correlate with reducing the risk of overuse injury. If so, then
bowlers whose body position at ball release has more reduced
redundancy than others should be at higher risk of injury, because
such bowlers are less able to adapt. We also assume that micro-injury
is more likely in fatigued tissues, and hence adopt the view that
reducing the probability of overuse injury is equivalent to reducing
fatigue. 

\subsection{The role of the various joint axes}

We interpret a joint axis that does not belong to the support in a 
given body position as a ``necessary'' axis of that position.

The joint axis $X$ is through the ``waist'' joint and perpendicular to
the pelvis; it is more or less parallel to the direction of the ball
around the time of release. It is always a  necessary axis, so for a
particular desired motion, the amount of sideways bending of the spine
is prescribed\footnote{Some care is needed here: in our model, bending
of the spine is approximated by rotation of an inflexible spine in the
``waist'' joint. It may be that more than one pattern of rotations of
vertebrae can achieve the desired rotation.}. One implication of this
is that injury risk due to this motion cannot be modified.

We noted above that a largely supported redundancy should help to
reduce fatigue. Similarly, if an axis is necessary then no fatigue
management can reduce the rate of tiring in structures involved in
rotations around it.  While the human body will have many more joint
axes, our analysis  suggests that bowlers will find it hard to
compensate for fatigue related to  rotation around the
$X$-axis. Anecdotal evidence suggests that bowlers may attempt
compensation by ``falling over'' as they tire. However, studies on
changes  in bowling action over long spells~\cite{glazetal00:1013}
have not reported rotation around this axis, so no scientific
judgement is possible. 

In critical positions we even suffer from reduced redundancy. The
calculations for reduced redundancy depend an the choice of shoulder
axes. We argued above that the $s_1$ axis is anatomically an intrinsic
axis of rotation, and that $s_2$ is dynamically an intrinsic axis of
rotation for fast bowlers, because the motion of the arm is in the
$s_1s_3$ plane around the time of delivery of a fast ball.

In the bowling of a cricket fast ball, the worst case scenario of reduced
redundancy is that $% 
S_{1}$ passes through the waist joint, which corresponds to case~3 above,
and implies that rotation about the other two shoulder axes are
prescribed in all motions. Let us consider the simplest (and also most
common) example: a straight arm. For such bowlers, the most risky
action is one in which wrist, elbow, shoulder and waist all lie on the
same line very near or at the moment of delivery. Their ability to
modify the amount of rotation will be limited to axes $S_1$, $Y$ and
$Z$; thus one expects overuse injury related to such rotations to be
less common. So for them trade-offs are only possible among axial
rotations of the arm, twisting of the spine, and bending forward at
the waist. On the other hand, coaches need to be aware that changing
the rotation in one of these axes will cause compensation in the other
two axes. 

We note that this situation is avoided by releasing the ball either behind
or in front of the plane of the torso (more on this below), by
a round-arm action, where $S_{1}$ is nearer to horizontal, and by a
very upright action, where $S_{1}$ is nearer to vertical. Vertical
action is usually encouraged by coaches, but in some cases may tend to align
the wrist with shoulder and waist, and so increase injury risk.

Furthermore, $Y$ and $Z$ are never necessary, so that the amount of twisting
and bending (backwards/forwards, that is) can be modified. So in case of
excessive rotation in these directions at the waist, it should be possible
to modify the bowler's action to reduce these, no matter what the
configuration of their joints at the moment of delivery. For instance,
excessive twisting around the $Z$-axis during the delivery stride is currently 
regarded as a major source of injury risk (the ``mixed'' action, which starts
with hips and shoulders facing forwards, then the shoulders rapidly rotate and 
counter-rotate, see~\cite{glazetal00:1013,mcgrfinc96} and many
others). Our study  suggests that bowlers using a mixed action should
be able to change action with  relative ease. 

Finally, is it possible to deliver a cricket ball with a maximally supported
redundancy? Yes, but such actions are unusual and discouraged by
coaches. The Line  $\mathbf{L}$ in the analysis above is the line
through waist and shoulder; it is required that this line is
perpendicular to none of the shoulder joint axes. For instance,
suppose that at delivery, the $X$ and $S_3$ axes are parallel
(certainly an aim in some deliveries by fast and medium pace
bowlers).  Then the wrist should not be in the plane formed by spine
and shoulder (otherwise case~2 applies: $\mathbf{L}$ perpendicular  
to $S_2$ or equivalently, $\mathbf{L}$ is a linear combination of 
$\mathbf{S_{1}}$ and $\mathbf{S_{3}}$).  So these bowlers should 
deliver such balls from behind or in front of the torso (the former
seems to be common). The other axes have similar
requirements. $\mathbf{L}$ perpendicular to  $S_{1}$ would be an excessively
round-arm action and perhaps unlikely (though it could occur in 
the slinging action of some fast bowlers). $\mathbf{L}$ perpendicular
to $S_{3}$ is perhaps harder to avoid but  
should still be rare; for instance a round-arm action with the
palm down at the moment of release, which might occur in some spin bowling 
actions.

\subsection{An example}

We give an analysis of the action of two medium-fast bowlers, both
from the youth academy and hence at risk of injury, as potentially
elite medium-fast or fast bowlers. The data kindly provided by Janine Gray
of Sports Science Institute of South Africa, who collected the data on
these two bowlers as part of a larger study. Both subjects were
17~years old and free of injury at the time the 
data were collected. Bowler~B had a long history of injuries, some of
them from non-cricket activities. In particular, he had suffered a
stress injury to the lower back, which was seen as due to
cricket. Bowler~A had never been injured. 
Their historical workloads were different---Bowler~B had played
cricket from early boyhood, while Bowler~A was a recent recruit into
the game.

\begin{figure}
\includegraphics[scale=.8]{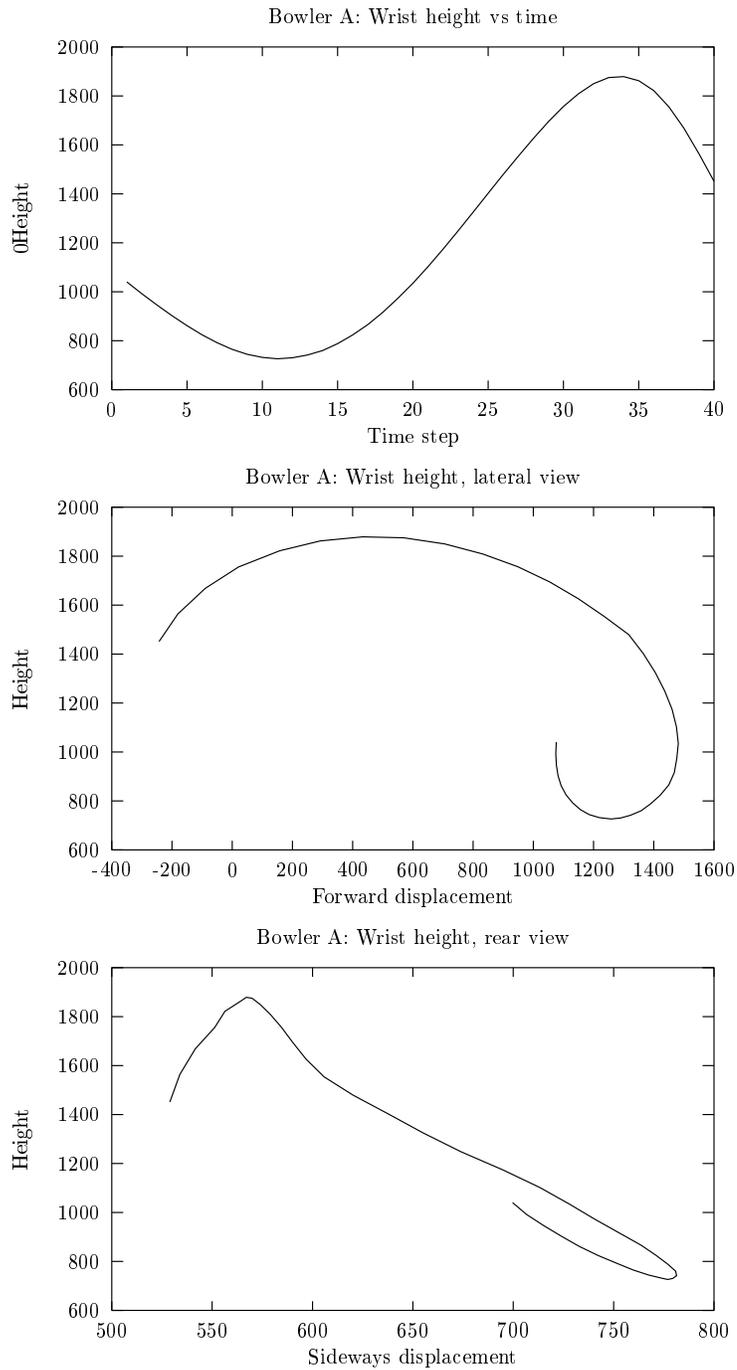}
\caption{Wrist position as a function of time for bowler A. The three
diagrams give different aspects. A) Wrist height as a function of time
B) Wrist path in the $YZ$-plane (movement is leftwards on diagram)
  C) Wrist path in the $XZ$-plane. Height and displacement in
  millimetres; time steps 0.83~milliseconds apart.}
\label{rawdataA}
\end{figure}

For each bowler, reflectors were attached to the body surface. Under
stroboscopic lighting (frequency 120~Hz), video cameras recorded the
positions of the reflectors at intervals 
(interval length about 8~ms). The following reflectors were used in
the calculation below: two on the wrist,  one on the shoulder 
and three on the waist. The three waist coordinates were assumed to lie
at the vertices of a symmetric trapezium, and the centre of its
circumrectangle  
was calculated to give $w$, the centre of the waist joint. In
calculating $s$, the centre of the shoulder joint, we assumed that the
shoulder is fixed  relative to the waist, so a simple correction
allowed us to move from the position of the reflector on the acromion
to $s$. The mid-point  of the two reflectors on the wrist provided the
position of $r$, the centre of the wrist. In  figures~\ref{rawdataA}
and~\ref{rawdataA} we depict aspects of the raw data:  wrist position
as a function of time for  both bowlers.

\begin{figure}
\includegraphics[scale=.8]{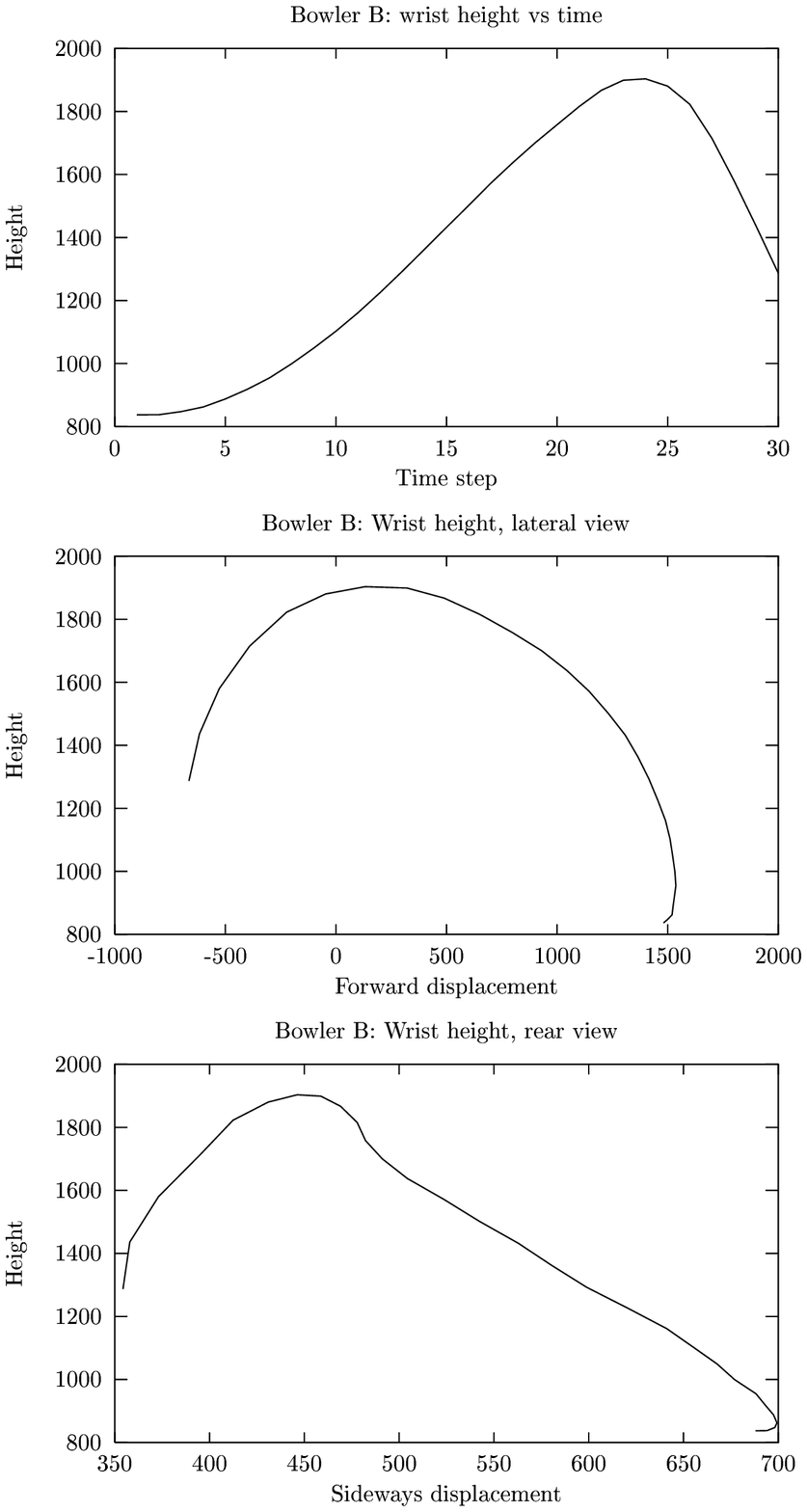}
\caption{Wrist position as a function of time for bowler B. The three
diagrams give different aspects. A) Wrist height as a function of time
B) Wrist path in the $YZ$-plane  C) Wrist path in the
$XZ$-plane. Height and displacement in 
  millimetres; time steps 0.83~milliseconds apart.}
\label{rawdataB}
\end{figure}

\begin{figure}
\includegraphics{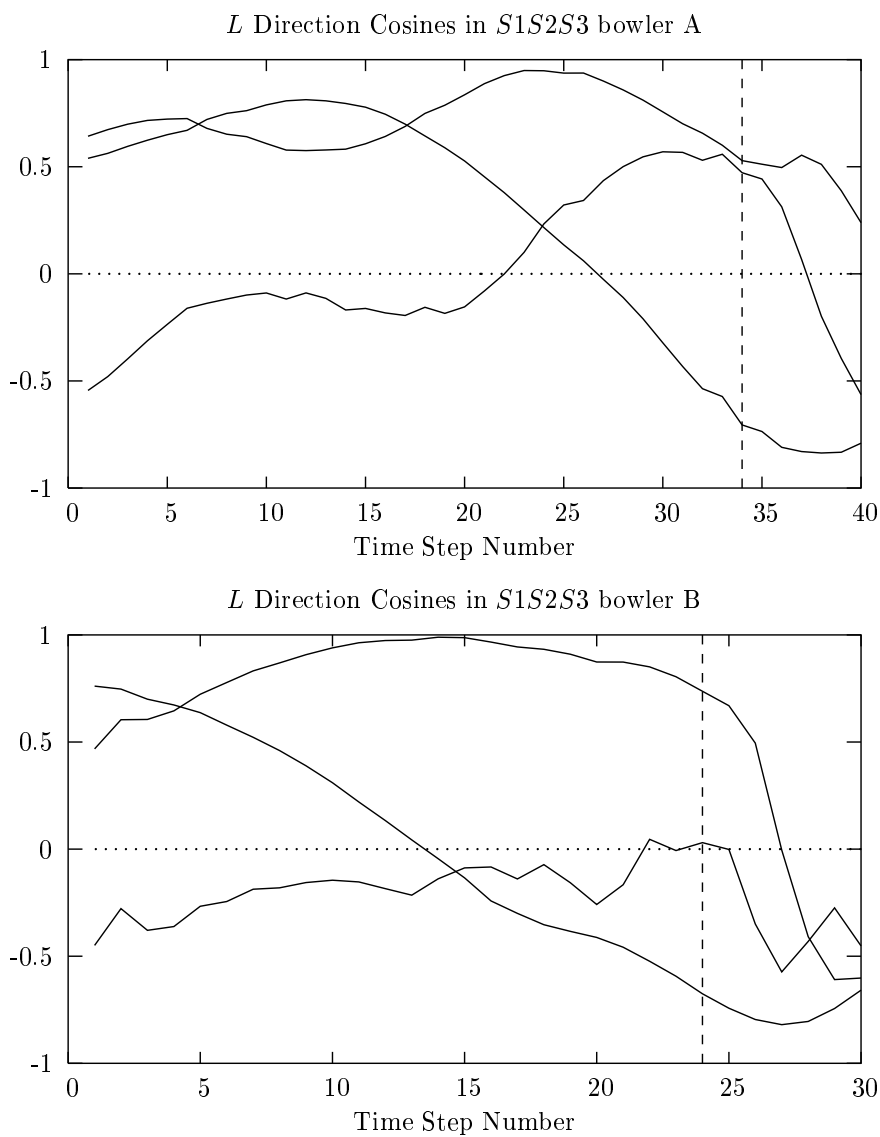}
\caption{Reduced redundancy as revealed by the direction of 
the waist-shoulder line in shoulder coordinates. Curves give 
direction cosines of $L$, the line from waist to shoulder, in 
the axes $S1$, $S2$ and $S3$ of the shoulder; they are plotted against
time steps (one unit of time is a few milliseconds). Vertical lines 
indicate
the approximate points of release. Whenever one of the direction
cosines goes to zero, reduction of redundancy occurs and the corresponding 
axis is absent from the support of the body 
position. Bowler A maintains full support until well after the moment
of delivery, but bowler B loses the $S_3$ axis from the support of the motion
for about 15~ms on either side of the moment of release.}
\label{dircossamples}
\end{figure}

Calculation of redundancy then proceeds as follows. The simple
subtraction $w-s$ and normalisation gave us  the unit vector
$\mathbf{l}$, which gives the direction of the Line $\mathbf{L}$
through waist and shoulder, and similarly $s-r$  gives $\mathbf{s_1}$,
the direction of $S_1$. Since the wrist reflectors lie in the
$S_1S_2$-plane as does $\mathbf{s_1}$, simple orthogonalisation gives
$\mathbf{s_2}$, and $\mathbf{s_3}$ is then available as the cross
product of  $\mathbf{s_1}$ and $\mathbf{s_2}$. The dot products of
$\mathbf{l}$  with the $\mathbf{s_i}$  are then calculated, giving the
direction  cosines of $\mathbf{l}$ in the $S_1S_2S_3$ axes. When
reduced redundancy  occurs, then $\mathbf{l}$ is perpendicular to one
or two of the  $\mathbf{s_i}$---that is, one of direction cosines are
zero. This is  easily spotted on a graph of direction cosines vs time (see 
Figure~\ref{dircossamples}).

The plots in Figure~\ref{dircossamples} cover approximately 0.4~s in
time and forward motion of about 2~m in  space. Note that for bowler~A
all three direction cosines stay well away from zero in the period
prior to release, but that for bowler~B the $S_3$ axis goes to zero
about 15~ms before release, and stays there for about 30~ms. With
respect to our choice of axes, bowler~B operates with reduced
redundancy around the time of release of the ball, but not
bowler~A. This suggests that bowler~B may be less able to modify his
action to cope with fatigue. This is consistent with their injury
history, as bowler~B indeed has had more injuries than
bowler~A. However, it is also true that Bowler~B has had much more
opportunity to for overuse than Bowler~A, due to a far longer playing
career. We hope to track both subjects to test whether indeed Bowler~A
will remain relatively free from overuse injury, as we predict.

\clearpage

\bibliography{../lprefs}

\end{document}